\documentstyle[preprint,epsfig,eqsecnum,aps,floats,tighten]{revtex}

\def\d0{D\O}
\def\D0{D\O}

\def\etmisv {\mbox{${\hbox{${\vec E}$\kern-0.6em\lower-.1ex\hbox{/}}}_T$}}
\def\etmis  {\mbox{${\hbox{$E$\kern-0.6em\lower-.1ex\hbox{/}}}_T$}}

\def\ifmath#1{\relax\ifmmode #1\else $#1$}%
\def\TeV{\ifmmode {\mathrm{ Te\kern -0.1em V}}\else
                   \textrm{Te\kern -0.1em V}\fi}%
\def\GeV{\ifmmode {\mathrm{ Ge\kern -0.1em V}}\else
                   \textrm{Ge\kern -0.1em V}\fi}%
\def\MeV{\ifmmode {\mathrm{ Me\kern -0.1em V}}\else
                   \textrm{Me\kern -0.1em V}\fi}%
\def\GeVcc{\ifmmode {\mathrm{ \GeV/c^2}}\else
                   \textrm{Ge\kern -0.1em V/c$^2$}\fi}%
\def\MeVcc{\ifmmode {\mathrm{ \MeV/c^2}}\else
                   \textrm{Me\kern -0.1em V/c$^2$}\fi}%

\def\Aslash{\mbox{${\hbox{$A$\kern-0.55em\hbox{/}}}$}}
\def\pslash{\mbox{${\hbox{$p$\kern-0.45em\hbox{/}}}$}}

\def\to{\rightarrow}

\def\gesim{\,{\raise-3pt\hbox{$\sim$}}\!\!\!\!\!{\raise2pt\hbox{$>$}}\,}
\def\lesim{\,{\raise-3pt\hbox{$\sim$}}\!\!\!\!\!{\raise2pt\hbox{$<$}}\,}
\def\boldoverdot{\,{\raise6pt\hbox{\bf.}\!\!\!\!\>}}

\def\diag{\hbox{\diag}}

\def\doubleundertext#1{
{\undertext{\vphantom{y}#1}}\par\nobreak\vskip-\the\baselineskip\vskip4pt%
\undertext{\hbox to 2in{}}}
\def\inbox#1{\vbox{\hrule\hbox{\vrule\kern5pt
     \vbox{\kern5pt#1\kern5pt}\kern5pt\vrule}\hrule}}
\def\sqr#1#2{{\vcenter{\hrule height.#2pt
      \hbox{\vrule width.#2pt height#1pt \kern#1pt
         \vrule width.#2pt}
      \hrule height.#2pt}}}

\def\today{\ifcase\month\or
  January\or February\or March\or April\or May\or June\or
  July\or August\or September\or October\or November\or December\fi
  \space\number\day, \number\year}
\def\pmb#1{\setbox0=\hbox{#1}%
  \kern-.025em\copy0\kern-\wd0
  \kern.05em\copy0\kern-\wd0
  \kern-.025em\raise.0433em\box0 }
\def\sumprime_#1{\setbox0=\hbox{$\scriptstyle{#1}$}
  \setbox2=\hbox{$\displaystyle{\sum}$}
  \setbox4=\hbox{${}'\mathsurround=0pt$}
  \dimen0=.5\wd0 \advance\dimen0 by-.5\wd2
  \ifdim\dimen0>0pt
  \ifdim\dimen0>\wd4 \kern\wd4 \else\kern\dimen0\fi\fi
\mathop{{\sum}'}_{\kern-\wd4 #1}}
\def\edge{$\widetilde{\mathrm{C}}$}

\begin{document}


\title{\vspace{1cm} Improved \d0\ W Boson Mass Determination}
\author{\centerline{The D\O\ Collaboration
\thanks{Submitted to the {\it International Europhysics Conference
        on High Energy Physics}, July 12--18, 2001, Budapest, Hungary, 
	and to the {\it XXth International Symposium on Lepton and Photon
        Interactions at High Energies}, July 23--28, 2001, Rome, Italy. 
	 }}}

\address{
\centerline{Fermi National Accelerator Laboratory, Batavia, Illinois 60510}
}

\date{\today}

\maketitle

\begin{abstract}
We present a measurement of the $W$ boson mass in proton-antiproton
collisions at $\sqrt{s} = 1.8$~TeV based on a data sample of
82~pb$^{-1}$ integrated luminosity collected by the \d0\ detector at
the Fermilab Tevatron. We utilize $e \nu$ events in which the electron
shower is close to the phi edge of one of the 32 modules in the \d0\
central calorimeter. The electromagnetic calorimenter response and
resolution in this region differs from that in the rest of the module
and electrons in this region were not previously utilized. We
determine the calorimeter response and resolution in this region using
$Z \to ee$ events.  We extract the $W$ boson mass by fitting to the
transverse mass and to the electron and neutrino transverse momentum
distributions. The result is combined with previous \d0\ results to
obtain an improved measurement of the $W$ boson mass: $m_W = 80.483
\pm 0.084$~GeV.
\end{abstract}

\newpage
\begin{center}
%
V.M.~Abazov,$^{23}$                                                           
B.~Abbott,$^{58}$                                                             
A.~Abdesselam,$^{11}$                                                         
M.~Abolins,$^{51}$                                                            
V.~Abramov,$^{26}$                                                            
B.S.~Acharya,$^{17}$                                                          
D.L.~Adams,$^{60}$                                                            
M.~Adams,$^{38}$                                                              
S.N.~Ahmed,$^{21}$                                                            
G.D.~Alexeev,$^{23}$
% ...
\\                                                                            
\vskip 0.30cm                                                                 
\centerline{(D\O\ Collaboration)}                                             
\vskip 0.30cm                                                                 
\centerline{$^{1}$Universidad de Buenos Aires, Buenos Aires, Argentina}       
\centerline{$^{2}$LAFEX, Centro Brasileiro de Pesquisas F{\'\i}sicas,         
                  Rio de Janeiro, Brazil}                                     
\centerline{$^{3}$Universidade do Estado do Rio de Janeiro,                   
                  Rio de Janeiro, Brazil}                                     
\centerline{$^{4}$Institute of High Energy Physics, Beijing,                  
                  People's Republic of China}                                 
\centerline{$^{5}$Universidad de los Andes, Bogot\'{a}, Colombia}             
\centerline{$^{6}$Charles University, Center for Particle Physics,            
                  Prague, Czech Republic}                                     
\centerline{$^{7}$Institute of Physics, Academy of Sciences, Center           
                  for Particle Physics, Prague, Czech Republic}               
\centerline{$^{8}$Universidad San Francisco de Quito, Quito, Ecuador}         
\centerline{$^{9}$Institut des Sciences Nucl\'eaires, IN2P3-CNRS,             
                  Universite de Grenoble 1, Grenoble, France}                 
\centerline{$^{10}$CPPM, IN2P3-CNRS, Universit\'e de la M\'editerran\'ee,     
                  Marseille, France}                                          
\centerline{$^{11}$Laboratoire de l'Acc\'el\'erateur Lin\'eaire,              
                  IN2P3-CNRS, Orsay, France}                                  
\centerline{$^{12}$LPNHE, Universit\'es Paris VI and VII, IN2P3-CNRS,         
                  Paris, France}                                              
\centerline{$^{13}$DAPNIA/Service de Physique des Particules, CEA, Saclay,    
                  France}                                                     
\centerline{$^{14}$Universit{\"a}t Mainz, Institut f{\"u}r Physik,            
                  Mainz, Germany}                                             
\centerline{$^{15}$Panjab University, Chandigarh, India}                      
\centerline{$^{16}$Delhi University, Delhi, India}                            
\centerline{$^{17}$Tata Institute of Fundamental Research, Mumbai, India}     
\centerline{$^{18}$Seoul National University, Seoul, Korea}                   
\centerline{$^{19}$CINVESTAV, Mexico City, Mexico}                            
\centerline{$^{20}$FOM-Institute NIKHEF and University of                     
                  Amsterdam/NIKHEF, Amsterdam, The Netherlands}               
\centerline{$^{21}$University of Nijmegen/NIKHEF, Nijmegen, The               
                  Netherlands}                                                
\centerline{$^{22}$Institute of Nuclear Physics, Krak\'ow, Poland}            
\centerline{$^{23}$Joint Institute for Nuclear Research, Dubna, Russia}       
\centerline{$^{24}$Institute for Theoretical and Experimental Physics,        
                   Moscow, Russia}                                            
\centerline{$^{25}$Moscow State University, Moscow, Russia}                   
\centerline{$^{26}$Institute for High Energy Physics, Protvino, Russia}       
\centerline{$^{27}$Lancaster University, Lancaster, United Kingdom}           
\centerline{$^{28}$Imperial College, London, United Kingdom}                  
\centerline{$^{29}$University of Arizona, Tucson, Arizona 85721}              
\centerline{$^{30}$Lawrence Berkeley National Laboratory and University of    
                  California, Berkeley, California 94720}                     
\centerline{$^{31}$University of California, Davis, California 95616}         
\centerline{$^{32}$California State University, Fresno, California 93740}     
\centerline{$^{33}$University of California, Irvine, California 92697}        
\centerline{$^{34}$University of California, Riverside, California 92521}     
\centerline{$^{35}$Florida State University, Tallahassee, Florida 32306}      
\centerline{$^{36}$University of Hawaii, Honolulu, Hawaii 96822}              
\centerline{$^{37}$Fermi National Accelerator Laboratory, Batavia,            
                   Illinois 60510}                                            
\centerline{$^{38}$University of Illinois at Chicago, Chicago,                
                   Illinois 60607}                                            
\centerline{$^{39}$Northern Illinois University, DeKalb, Illinois 60115}      
\centerline{$^{40}$Northwestern University, Evanston, Illinois 60208}         
\centerline{$^{41}$Indiana University, Bloomington, Indiana 47405}            
\centerline{$^{42}$University of Notre Dame, Notre Dame, Indiana 46556}       
\centerline{$^{43}$Iowa State University, Ames, Iowa 50011}                   
\centerline{$^{44}$University of Kansas, Lawrence, Kansas 66045}              
\centerline{$^{45}$Kansas State University, Manhattan, Kansas 66506}          
\centerline{$^{46}$Louisiana Tech University, Ruston, Louisiana 71272}        
\centerline{$^{47}$University of Maryland, College Park, Maryland 20742}      
\centerline{$^{48}$Boston University, Boston, Massachusetts 02215}            
\centerline{$^{49}$Northeastern University, Boston, Massachusetts 02115}      
\centerline{$^{50}$University of Michigan, Ann Arbor, Michigan 48109}         
\centerline{$^{51}$Michigan State University, East Lansing, Michigan 48824}   
\centerline{$^{52}$University of Nebraska, Lincoln, Nebraska 68588}           
\centerline{$^{53}$Columbia University, New York, New York 10027}             
\centerline{$^{54}$University of Rochester, Rochester, New York 14627}        
\centerline{$^{55}$State University of New York, Stony Brook,                 
                   New York 11794}                                            
\centerline{$^{56}$Brookhaven National Laboratory, Upton, New York 11973}     
\centerline{$^{57}$Langston University, Langston, Oklahoma 73050}             
\centerline{$^{58}$University of Oklahoma, Norman, Oklahoma 73019}            
\centerline{$^{59}$Brown University, Providence, Rhode Island 02912}          
\centerline{$^{60}$University of Texas, Arlington, Texas 76019}               
\centerline{$^{61}$Texas A\&M University, College Station, Texas 77843}       
\centerline{$^{62}$Rice University, Houston, Texas 77005}                     
\centerline{$^{63}$University of Virginia, Charlottesville, Virginia 22901}   
\centerline{$^{64}$University of Washington, Seattle, Washington 98195}       

\end{center}

\normalsize


\section{Introduction}

Measurements of the $W$ boson mass are of fundamental interest since
they help constrain the Standard Model and the Higgs boson
mass\cite{erler_mhiggs}. Recent measurements of the $W$ boson mass
have been made by \d0~\cite{d0_mw} and CDF~\cite{cdf_mw} at the
Tevatron and by the LEP experiments~\cite{lep_mw}.  Previous
measurements by \d0\ have relied on $W
\to e \nu$ events in which the electron was detected in the central
calorimeter or the forward calorimeters.  The central calorimeter is
divided azimuthally into 32 modules~\cite{d0_nim}. Electrons incident
close to the azimuthal module boundaries were found to have a reduced
response and degraded energy resolution. These ``edge'' electrons were
excluded from our $W$ and $Z$ boson data samples in previous
measurements. In this paper, we report a new measurement of the $W$
boson mass using these edge electrons.  We rely on $Z \to ee $ events
in which at least one electron is detected in the edge region to
calibrate the response of the calorimeter. $Z \to e e$ events in which
only one electron is incident at a central calorimeter module boundary
are also used to additionally constrain the
elecromagnetic calorimeter enegry scale for non-edge electrons,
thereby improving our previous measurements based on non-edge
electrons.

\section{Event Selection and Detector Calibration}

Direct measurement of the $W$ boson mass $m_W$ at \d0\ is performed
using $W \to e \nu$ events from $p \bar p$ collisions at a
center-of-mass energy of 1.8~TeV.  A detailed description of the
method used to measure $m_W$ is given in Ref.~\cite{d0_mw}.
Events are selected by requiring the presence of an
isolated electron with high transverse momentum ($p_T$) and large
missing transverse energy ($\etmis$). The $W$ boson mass is extracted
by fitting Monte Carlo templates to the observed kinematic
distributions. Maximum likelihood fits are made to the transverse mass
$m_T = \sqrt{2 p_T^e p_T^\nu ( 1 - \mathrm{cos} \,
\phi_{e \nu})}$, electron transverse momentum $p_T^e$, and neutrino
transverse momentum $p_T^\nu$. Here, $\phi_{e \nu}$ is the azimuthal
angle between the electron and neutrino.  The three $W$ boson mass
measurements are combined taking into account correlations to obtain
the final result. A Monte Carlo simulation is used to provide the
expected lineshapes of the distributions as a function of $m_W$.  The
Monte Carlo contains all resolution effects and backgrounds as
determined from data.

The $W$ boson sample for this measurement is selected requiring
$\etmis > 25$~GeV and a high-quality isolated electron in the central
calorimeter (CC) with $p_T^e > 25$~GeV and $\Delta \phi < 0.1 \times
2\pi/32 = 0.02$~radians, where $\Delta \phi$ is the angle between the
electron direction and the closest CC module boundary. The electron
direction is calculated from the center-of-gravity of the track in the
central drift chamber and the event vertex position. Electrons
satisfying these criteria are referred to as ``\edge\ electrons'',
while non-edge electrons which have $\Delta \phi > 0.02$~radians are
called ``C electrons''. The number of candidate edge-electron $W$
events selected by applying the above criteria was $3\,853$. For
comparison, our previous central calorimeter measurement using the
1994-95 data set was based on $28\,323$ candidates.

We also select $Z \to ee$ candidates requiring two isolated electrons
with $p_T^e > 25$~GeV with dielectron invariant mass $60~{\mathrm GeV}
< m_{ee} < 120$~GeV.  Events are required to have one electron in the
edge region. The second electron may also be in the edge region
(\edge-\edge\ events), or it may be in the non-edege region (\edge-C
events), or in one of the end calorimeters (\edge-E events).  The
numbers of $Z$ candidates selected are 470 \edge-C events, 47
\edge-\edge events, and 154 \edge-E events. Backgrounds to the edge
electron $W$ and $Z$ samples are determined using the same methods
used in our previous analyses.

%

\begin{figure}[ht]
   \epsfysize = 9cm
   \centerline{\epsffile{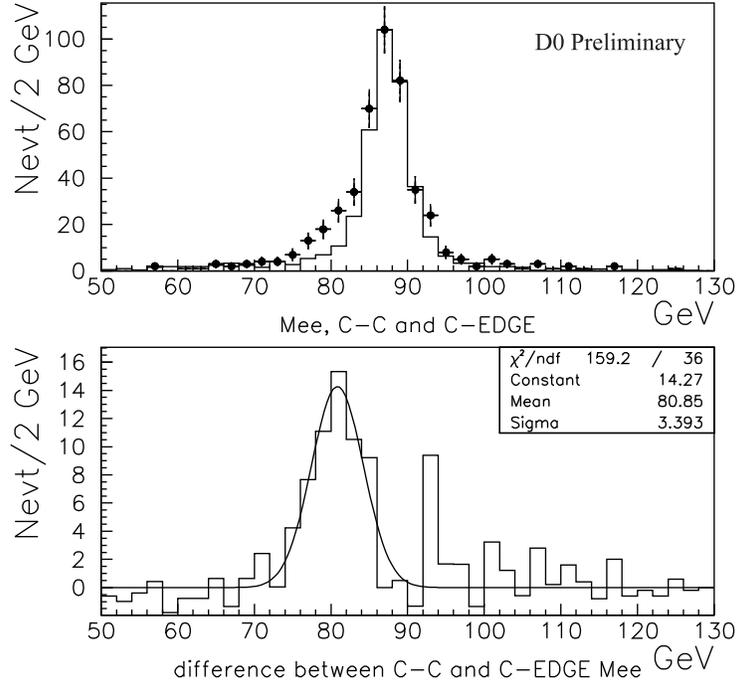}}
	\vspace{4mm}
   \caption{(a) Dielectron mass distribution for the edge sample
   (points with error bars) and the non-edge sample (solid line). (b)
   Difference between the two distributions in (a) fitted with a
   single Gaussian function.} 
\label{fig:mee_comp}
\end{figure}

The calorimeter response to edge electrons is illustrated in
Fig.~\ref{fig:mee_comp}, which compares the reconstructed dielectron
invariant mass distributions of \edge-C and C-C events.
Above the $Z$ peak, the distributions are consistent with one
another, but at low $m_{ee}$ there is an excess of events in the edge
sample indicating that a fraction of the edge electrons have a
lower electromagnetic response in the calorimeter.
The difference between the distributions is well described by a single
Gaussian function. This suggests that the electromagnetic calorimeter
response for edge-electrons can be described by the sum of two
Gaussians, one with the same mean and width as for non-edge electrons
and the second with a reduced response and degraded energy resolution.
This is consistent with expectations, since the high voltage
electrodes are set back near the module edge, thus reducing the
electric field in that region and giving lower response.  There is no
evidence for increased energy deposit in the backing hadron
calorimeter module that would occur if particles were passing within a
crack between EM modules. We assume that a fraction $f_{\mathrm
edge}$ of the edge electrons has a reduced response and degraded
energy resolution, while the remaining edge electrons have the same
response and energy resolution as non-edge electrons. Thus, for the
fraction $f_{\mathrm edge}$ of edge electrons, the calorimeter
response is parameterized by
\begin{eqnarray*}
E^{\mathrm meas} =
\alpha_{\mathrm edge} E^{\mathrm true}_e + \delta
\end{eqnarray*}

\noindent
The offset $\delta$ was found to be consistent with the offset 
previously used in the parameterization of non-edge electrons, while the scale
$\alpha_{\mathrm edge}$ must be separately determined for the edge
electrons.  The energy resolution is parameterized by:
\begin{eqnarray*}
\left( \frac{\sigma_{E}}{E} \right)^2 =  ( c_{\mathrm edge} )^2 
+ \left( \frac{s}{\sqrt{E}} \right)^2
+ \left( \frac{n}{E} \right)^2
\end{eqnarray*}

\noindent
where the sampling term $s$ and noise term $n$ are the same as for
non-edge electrons. The parameters $f_{\mathrm edge}$,
$\alpha_{\mathrm edge}$, and $c_{\mathrm edge}$ are determined by
fitting the invariant mass distribution of \edge-C events to two
Gaussians, assuming a $Z$ boson mass equal to the measured LEP value.
This fit gives
\begin{eqnarray*}
f_{\mathrm edge} &=& 0.346 \pm 0.076 \\
\alpha_{\mathrm edge}&=& 0.912 \pm 0.018 \\
c_{\mathrm edge} &=& 0.101^{+0.028}_{-0.018}.
\end{eqnarray*}

\begin{figure}[ht]
   \epsfysize = 9cm
   \centerline{\epsffile{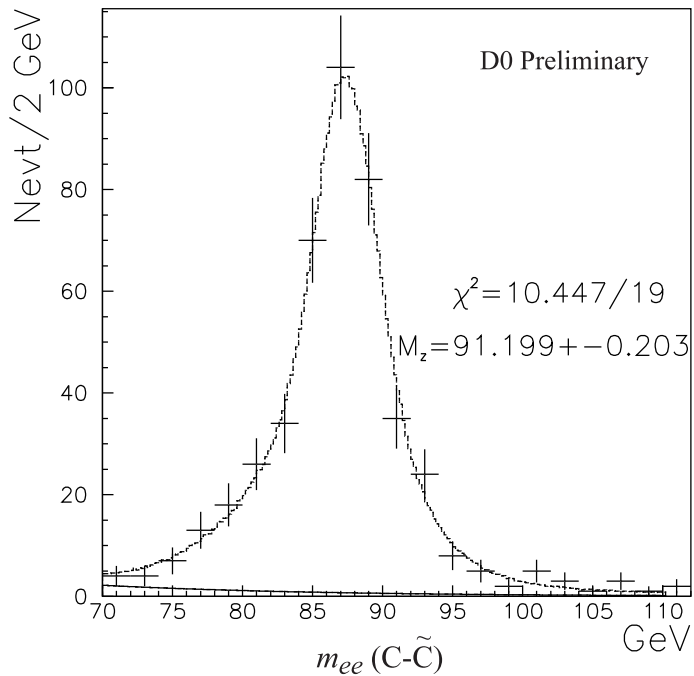}}
	\vspace{3mm}
   \caption{Dielectron mass distribution for \edge-C events. The dashed 
            histogram shows the maximum likelihood fit and the solid curve 
            is the background contribution.} 
\label{fig:mee_fit}
\end{figure}

Figure~\ref{fig:mee_fit} shows a fit to the dielectron invariant mass
distribution using the sum of two Gaussians, one with the edge
parameters determined above and the other with the parameters for
non-edge electrons previously determined from C-C events. The
parameterization gives a good description of the observed data.

\section{Results}

\begin{figure}[hp]
   \epsfysize = 7cm
   \centerline{\epsffile{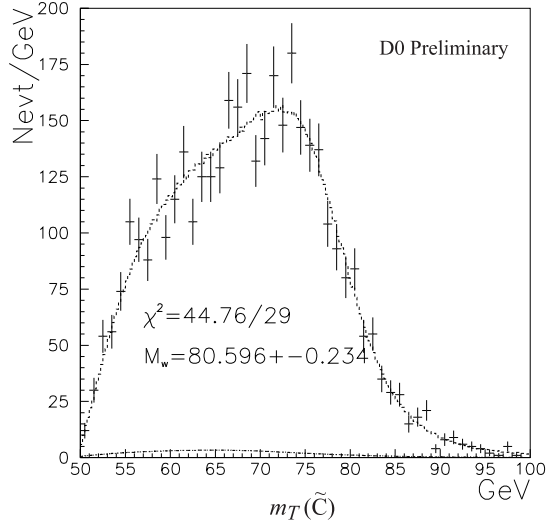}}
	\vspace{2mm}
   \centerline{\epsffile{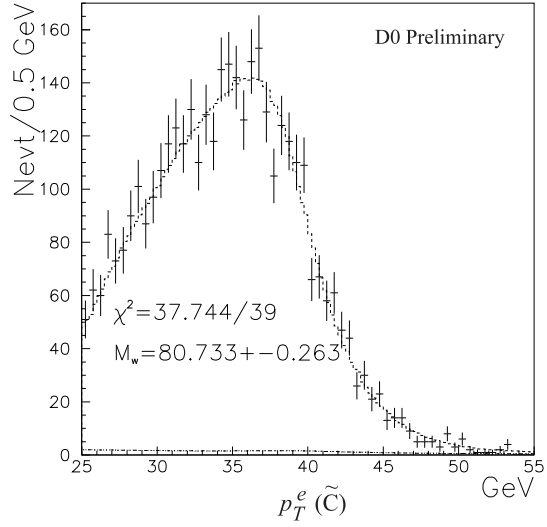}}
	\vspace{2mm}
   \centerline{\epsffile{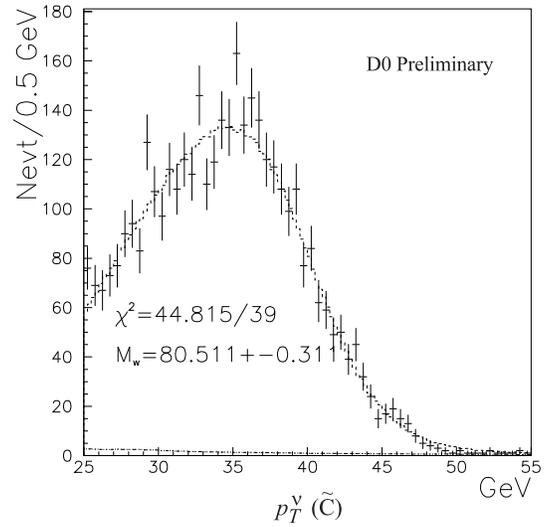}}
	\vspace{3mm}
   \caption{Distributions of $m_T$, $p_T^e$, and $p_T^\nu$ from the edge 
            electron $W$ data. The superimposed dashed histograms show the
            maximum likelihood fits and the solid curves show the estimated
	    backgrounds.} 
\label{fig:fits}
\end{figure}

The results of the fits to the transverse mass and electron and
neutrino transverse momentum distributions are shown in
Fig.~\ref{fig:fits}. 
The results are: 
\begin{eqnarray*}
m_W &=& 80.596 \pm 0.234~{\mathrm (stat)~GeV},~~~\chi^2 = 45/29
~~~(m_T~{\mathrm fit}) \\
m_W &=& 80.733 \pm 0.263~{\mathrm (stat)~GeV},~~~\chi^2 = 
38/39~~~(p_T^e~{\mathrm fit}) \\
m_W &=& 80.511 \pm 0.311~{\mathrm (stat)~GeV},~~~\chi^2 = 
45/39~~~(p_T^\nu~{\mathrm fit})
\end{eqnarray*}

\noindent
The errors are statistical only. The systematic errors are listed in
Table~\ref{table:syst_err}. Combining these measurements taking into
account systematic errors and their correlations gives the final
result for the edge electron $W$ mass:
\begin{eqnarray*}
m_W &=& 80.574 \pm 0.405~{\mathrm GeV}
\end{eqnarray*}

\begin{table}[h]
\begin{center}
\begin{minipage}{12cm}
\begin{tabular}{lrrr}
Source               & $m_T$ Fit & $p_T^e$ Fit & $p_T^\nu$ Fit \\ \hline
 $W$ Statistics                       &  234  &  263   & 311 \\ 
 Edge EM scale ($\alpha_{\mathrm edge}$)          &  265  &  309   & 346 \\ 
 CC EM scale  ($\alpha_{cc}$)     &  128  &  131   & 113 \\ 
 CC EM offset ($\delta_{cc}$)     &  142  &  139   & 145 \\
 Calorimeter uniformity           &  10   &  10    & 10  \\
 Electron angle calibration                        &  38   &  40    & 52  \\ 
 Backgrounds                      &  10   &  20    & 20  \\ 
 CC EM resolution ($c_{cc}$)     &  15   &  18    & 2   \\ 
 Edge EM resolution ($c_{\mathrm edge}$) &  268  &  344   & 404 \\ 
 Fraction of events ($f_{\mathrm edge}$)  &   8   &  14    & 22  \\ 
 Recoil response                &  20   &  16    & 46  \\ 
 Recoil resolution              &  25   &  10    & 90  \\ 
 Electron removal                &  15   &  15    & 20  \\ 
 Selection bias                &  2    &  9     & 20  \\ 
 Parton luminosity                &  9    &  11    & 9   \\ 
 Radiative corrections            &  4    &  8     & 0   \\ 
 PDF                              &  0    &  64    & 9   \\ 
 $p_T(W)$                & 10    &  50    & 25  \\ 
 $W$-boson width                        & 10    &  10    & 10  \\ 
\end{tabular}
\end{minipage}
\end{center}
\caption[tab1]{$W$ mass uncertainties (in MeV) in the edge electron measurements.
         The uncertainties due to the edge electron parameters 
         $f_{\mathrm edge}$, $\alpha_{\mathrm edge}$, and $c_{\mathrm edge}$ are
         explained in the text, while details of the other sources of 
         uncertainty are given in Ref.~\cite{d0_mw}.}
\label{table:syst_err}
\end{table}


The \edge-C $Z \to ee$ data sample provides a means to additionally
constrain the central calorimeter scale $\alpha_{CC}$ and resolution
constant term $c_{CC}$ for non-edge electrons. Fitting to the observed
$m_{ee}$ distribution yields $\alpha_{CC} = 0.9552 \pm 0.0023$. The
\edge-E events can also be used to fit for $\alpha_{CC}$ and
$\alpha_{EC}$ yielding $\alpha_{CC} = 0.9559 \pm 0.0107$ and
$\alpha_{EC} = 0.9539 \pm 0.0085$. These values are consistent with
the results obtained in our earlier analyses of non-edge and EC events
and can be combined with them taking into account the correlations to
improve the energy scale uncertainty, and hence the uncertainty on the
$W$ boson mass measurement.

\section{Combined W mass results}

To obtain the final result for the $W$ boson mass, we combine the following 
measurements:

\begin{itemize}
\item[(i)] The Run 1a $W$ mass measurement from a fit to $m_T$

\item[(ii)] The three Run 1b central calorimeter measurements from fits to $m_T$,
$p_T^e$, and $p_T^\nu$

\item[(iii)] The three Run 1b end calorimeter measurements from fits to $m_T$,
$p_T^e$, and $p_T^\nu$

\item[(iv)] The three edge electron measurements from fits to $m_T$,
$p_T^e$, and $p_T^\nu$

\end{itemize}

\noindent
The measurements in (ii) and (iii) include the improvement due to the
additional constraints on the EM calorimeter energy scale from edge
events as discussed above.
 
The final combined result is
\begin{eqnarray*}
m_W = 80.483 \pm 0.084~{\mathrm GeV}
\end{eqnarray*}

\noindent
This represents an improved error of 7~MeV over our previously
published result ($80.482 \pm 0.091$~GeV~\cite{d0_mw}). A major part
of the improved uncertainty is due to the use of the
\edge-C events to constrain the EM calorimeter energy scale for
non-edge electrons.

\section{Conclusion}

We have improved the uncertainty in the \d0\ measurement of the $W$
boson mass, using $W \to e \nu$ and $Z \to ee$ events in which
electrons are detected in the edge region at the boundary between
modules of the central calorimeter. The new result is $m_W = 80.483 \pm
0.084$~GeV.

\section*{Acknowledgements}
%
We thank the staffs at Fermilab and collaborating institutions, 
and acknowledge support from the 
Department of Energy and National Science Foundation (USA),  
Commissariat  \` a L'Energie Atomique and 
CNRS/Institut National de Physique Nucl\'eaire et 
de Physique des Particules (France), 
Ministry for Science and Technology and Ministry for Atomic 
   Energy (Russia),
CAPES and CNPq (Brazil),
Departments of Atomic Energy and Science and Education (India),
Colciencias (Colombia),
CONACyT (Mexico),
Ministry of Education and KOSEF (Korea),
CONICET and UBACyT (Argentina),
The Foundation for Fundamental Research on Matter (The Netherlands),
PPARC (United Kingdom),
Ministry of Education (Czech Republic),
and the A.P.~Sloan Foundation.
%

%
%

\end{document}